\documentclass[twocolumn,aps,prl,preprintnumbers]{revtex4-2}
\usepackage{bbm}
\usepackage[latin9]{inputenc}
\usepackage{float}
\usepackage{amsmath}
\usepackage{amssymb}
\usepackage{graphicx}
\usepackage{mathrsfs}
\usepackage{amsfonts}
\usepackage{amsthm}
\usepackage{color}
\usepackage{txfonts}
\usepackage[colorlinks=true,citecolor=blue,linkcolor=blue,urlcolor=blue,anchorcolor=blue]{hyperref}%
\usepackage{pifont}
\hypersetup{colorlinks=true,citecolor=blue,linkcolor=blue,urlcolor=blue}
\providecommand{\U}[1]{\protect\rule{.1in}{.1in}}
\makeatletter
\@ifundefined{textcolor}{}
{
\definecolor{BLACK}{gray}{0}
\definecolor{WHITE}{gray}{1}
\definecolor{RED}{rgb}{1,0,0}
\definecolor{GREEN}{rgb}{0,1,0}
\definecolor{BLUE}{rgb}{0,0,1}
\definecolor{CYAN}{cmyk}{1,0,0,0}
\definecolor{MAGENTA}{cmyk}{0,1,0,0}
\definecolor{YELLOW}{cmyk}{0,0,1,0}
}

\makeatother
\begin{document}
\title{Skyrmion Hall effect in altermagnets}
\author{Zhejunyu Jin}
\author{Zhaozhuo Zeng}
\author{Yunshan Cao}
\author{Peng Yan}
\email[Corresponding author: ]{yan@uestc.edu.cn}
\affiliation{School of Physics and State Key Laboratory of Electronic Thin Films and Integrated Devices, University of
Electronic Science and Technology of China, Chengdu 610054, China}

\begin{abstract}
It is widely believed that the skyrmion Hall effect is absent in antiferromagnets because of the vanishing topological charge. However, the Aharonov-Casher theory indicates the possibility of topological effects for neutral particles. In this work, we predict the skyrmion Hall effect in emerging altermagnets with zero net magnetization and zero skyrmion charge. We first show that the neutral skyrmion manifests as a magnetic quadrupole in altermagnets. We reveal a hidden gauge field from the magnetic quadrupole, which induces the skyrmion Hall effect when driven by spin transfer torque. Interestingly, we identify a sign change of the Hall angle when one swaps the anisotropic exchange couplings in altermagnets. Furthermore, we demonstrate that both the velocity and Hall angle of altermagnetic skyrmions sensitively depend on the current direction. Our findings real the critical role of magnetic quadrupole in driving the skyrmion Hall effect with vanishing charge, and pave the way to discovering new Hall effect of neutral quasiparticles beyond magnetic skyrmions.
\end{abstract}

\maketitle

\textit{Introduction.---}The Hall effect is one of the most important phenomena in condensed matter physic, which holds significant potential in manipulating the particle and wave transports, besides fundamental interests  \cite{Hall,Hall1,Jungwirth2002,Nagaosa2010,Klitzing1980,Klitzing1986,Novoselov2007,Chang2023,Bernevig2006,
Kato2004,Sinova2015,Neubauer2006,Hoogdalem2013,Onose2010,Sodemann2015,Ma2019,Du2021,Jin2023}. Beyond the realm of standard fermions and bosons, Hall effects have also been found in various quasiparticle systems \cite{Shibata2006,Gobel2019,Iwasaki1,Liu2022,Gobel2019,Kim2019}, with a representative one being the skyrmion Hall effect \cite{Nagaosa2013,Zhang2017,Kim2017,Reichhardt2022}. Magnetic skyrmion is a real-space topological spin texture \cite{Skyrme1962,Muhlbauer2009,Yu2010}, and is attracting extensive attention in spintronics due to its potential for computing and information processing. Under external drivings, skyrmions with a finite topological charge will experience a gyrotropic force, leading to the skyrmion Hall effect \cite{Jiang2017}. In antiferromagnets, skyrmions are composed of ferromagnetic components in different sublattices with compensated charges, and thus are neutral. It is commonly believed that there is no Hall effect for ``neutral skyrmions" \cite{Barker2016,Zhang2016,Legrand2020,Pham2024}.

In their pioneering work, Aharonov and Casher proposed that neutral particles with a finite magnetic moment can accumulate geometric phases, due to the presence of the vector potential \cite{Aharonov1984}.  Nevertheless, its manifestation in skyrmionic systems is yet to be explored. Recently, an emerging class of magnet dubbed altermagnet (ATM) was identified \cite{Hayami2019,Hayami2020,Smejkal1,Smejkal2,Smejkal3,Bai2023,Feng2024,McClarty2024}, which maintains zero macroscopic magnetization but breaks the Kramers' degeneracy. Spin textures in ATMs carry zero topological charge but meanwhile have finite local magnetic moments, resembling composite neutral particles considered in the Aharonov-Casher (AC) formalism. This feature suggests the possibility to discover the Hall-like motion of neutral skyrmions in altermagnets.

In this Letter, we aim to reveal the Hall effect of charge-free spin textures with finite local magnetic moments. To this end, we theoretically study the skymion dynamics driven by spin transfer torque (STT) in ATM, without loss of generality. We first show that the neutral ATM skyrmion manifests as a magnetic quadrupole. A hidden gauge field from the magnetic quadrupole is predicted, which results in the skyrmion Hall effect when driven by STT.  Based on the collective-coordinate method, we analytically derive the equation of motion for ATM skyrmions. We find that the skyrmion Hall angle sensitively depends on the current direction, due to the anisotropic nature of magnetic quadrupoles. Interestingly, we observe that the ATM skyrmion Hall effect is absent when skyrmions in two sublattices are of mirror symmetry with respect to the axis of the current flow. Our results highlight the key role of the magnetic quadrupole in generating the skyrmion Hall effect in ATM.

\begin{figure}
  \centering
  \includegraphics[width=0.49\textwidth]{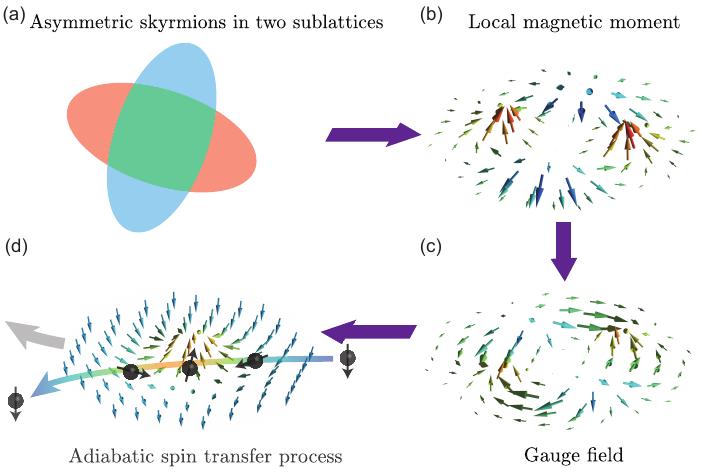}\\
  \caption{Schematic illustration of the ``neutral skyrmion" Hall effect in altermagnets. (a) Boundaries of red and blue ellipses denote the wall center ($s_z=0$) of orthogonal skyrmions in two sublattices. (b) Local magnetic moments in ATM skyrmion form a magnetic quadrupole that results an effective gauge field (c).  (d) Adiabatic spin transfer of electrons leads to the ATM skyrmion Hall effect (gray arrow).}\label{fig1}
\end{figure}

\textit{Model.---}Let's consider a two-sublattice ATM model with the following Hamiltonian \cite{alter}
\begin{equation}\label{Eq1}
\begin{aligned}
{\mathcal H}_{\text{ATM}}=&-\sum_{i,j}\big[J_{1}{\bf s}_{i,j}^A\cdot{\bf s}_{i+1,j}^A+J_{2}{\bf s}_{i,j}^B\cdot{\bf s}_{i+1,j}^B+J_{2}{\bf s}_{i,j}^A\cdot{\bf s}_{i,j+1}^A\\
&+J_{1}{\bf s}_{i,j}^B\cdot{\bf s}_{i,j+1}^B+J_3{\bf s}_{i,j}^A\cdot{\bf s}_{i,j}^B+D_0({\bf s}_{i,j}^A\times{\bf s}_{i+1,j}^A\\
&+{\bf s}_{i,j}^B\times{\bf s}_{i+1,j}^B)\cdot{\bf y}-D_0({\bf s}_{i,j}^A\times{\bf s}_{i,j+1}^A+{\bf s}_{i,j}^B\times{\bf s}_{i,j+1}^B)\cdot{\bf x}\\
&+K_0({\bf s}_{i,j}^A\cdot{\bf z})^2+K_0({\bf s}_{i,j}^B\cdot{\bf z})^2\big],\\
\end{aligned}
\end{equation}
where $J_{1,2}>0$ represents the intralayer ferromagnetic exchange strength, $J_3<0$ is the interlayer antiferromagnetic exchange constant, and $D_0$ and $K_0$ denote the interfacial Dzyaloshinskii-Moriya interaction (DMI) and magnetic anisotropy coefficients, respectively. ${\bf s}_{i,j}^A$ and ${\bf s}_{i,j}^B$ are the normalized spins on sites $(i,j)$ of sublattices $A$ and $B$, respectively. The anisotropic nature of Hamiltonian \eqref{Eq1} allows orthogonal elliptic skyrmions in two sublattices [see Fig. \ref{fig1}(a)]. To analytically study the dynamics of altermagnetic textures, we adopt the magnetic and N$\rm{\acute{e}}el$ order parameters ${\bf m}=({\bf s}_A+{\bf s}_B)/2$ and ${\bf l}=({\bf s}_A-{\bf s}_B)/2$ to rewrite the model in the continuous form (see Supplemental Material \cite{SM} for detailed derivations)
 \begin{equation}\label{Eq2}
\begin{aligned}
{\mathcal H}=&\int \Big [\frac{A_0}{2} {\bf m}^2+A_1 (\nabla {\bf l})^2+A_2(\partial_x {\bf m}\cdot\partial_x {\bf l}-\partial_y {\bf m}\cdot\partial_y {\bf l})\\
&+D l_z \nabla \cdot {\bf l}-D({\bf l}\cdot\nabla)n_z+K l_z^2 \Big ]d{\bf r}.\\
\end{aligned}
\end{equation}
The above Hamiltonian includes the homogeneous exchange, inhomogeneous exchange, anisotropic altemagnetic exchange, DMI, and magnetic anisotropy energies, with $A_0=-4J_3/d^3$, $A_1=(J_1+J_2)/d$, $A_2=(J_2-J_1)/d$, $D=2D_0/d^2$, and $K=2K_0/d^3$ being the coefficients, respectively. Here $d$ represents the lattice constant. It is noted that the altemagnetic exchange term $A_2(\partial_x {\bf m}\cdot\partial_x {\bf l}-\partial_y {\bf m}\cdot\partial_y {\bf l})$ breaks the symmetry of conventional antiferromagnets, i.e., Kramers' degeneracy. We can then write the Lagrangian
\begin{equation}\label{Eq3}
\mathcal{L}=\int \frac{1}{\gamma}(\partial_t {\bf l}\times {\bf l})\cdot {\bf m} d{\bf r}-{\mathcal H},\\
\end{equation}with $\gamma$ being the gyromagnetic ratio. By invoking the Euler-Lagrangian formula, the local magnetic moment $\bf m$ can be expressed as
\begin{equation}\label{Eq4}
\begin{aligned}
{\bf m}=\frac{1}{A_0}\Bigg \{\frac{1}{\gamma}\partial_t{\bf l}\times{\bf l}+A_2\Big [\partial_x^2{\bf l}-({\bf l}\cdot\partial_x^2{\bf l}){\bf l}-\partial_y^2{\bf l}+({\bf l}\cdot\partial_y^2{\bf l}){\bf l}\Big ] \Bigg \}.\\
\end{aligned}
\end{equation}
Equation \eqref{Eq4} allows us to eliminate ${\bf m}$ and to obtain an effective Lagrangian for the staggered field
\begin{equation}\label{Eq5}
\begin{aligned}
\mathcal{L}=\int  \Big [\frac{1}{A_0\gamma^2}(\partial_t {\bf l})^2 -U({\bf l})+\frac{1}{\gamma}{\bf \mathscr{A}}\cdot\partial_t{\bf l} \Big ]d{\bf r},\\
\end{aligned}
\end{equation}where the first term represents the kinetic energy of magnetic textures, the second term denotes the potential energy, and the last term is the gyrotropic term \cite{Ivanov1994,Dasgupta2017}, with the effective gauge field
\begin{equation}\label{Eq6}
\begin{aligned}
\mathscr{A}={\bf m}_{s}\times {\bf l}=\frac{A_2}{A_0} (\partial_x^2 {\bf l}-\partial_y^2 {\bf l})\times {\bf l}.\\
\end{aligned}
\end{equation}
Here, ${\bf m}_{s}={\bf m}-(\partial_t{\bf l}\times{\bf l})/(\gamma A_0)$ represents the static local magnetization density. It is analogous to the AC effect, which describes a magnetic moment moving in an electric field by comparing Eq. \eqref{Eq5} and  Eq. (9) in Ref. \cite{Aharonov1984}. Defining a map between the constituents of these equations, 
\begin{equation}\label{Mapping}
(\frac{1}{A_0}, \partial_t{\bf l}, {\bf m}_s,{\bf l}, \mathscr{A})\longleftrightarrow(M, {\bf v}, \boldsymbol{\mu}, {\bf E}, {\bf A}),
\end{equation} allows us to establish an isomorphism between the dynamics of the ATM skyrmion and AC effect. Here, $M$ is the mass of the magnetic moment $\boldsymbol{\mu}$ with a moving velocity ${\bf v}$, ${\bf E}$ is the external electric field, and ${\bf A}$ is the gauge field \cite{Aharonov1984}. However, the net magnetic moment in ATM is zero, i.e., $\langle{\bf m}_s\rangle=0$ by integrating over the ATM, which is slightly different from the case in the original AC formalism. Interestingly, we note that the local magnetic moments form a magnetic quadrupole that can be described by cubic harmonics \cite{SM,Hayami2018}, as shown in Fig. \ref{fig1}(b) and Fig. S2 \cite{SM}. Below, we derive the current-driven equation of motion for ATM skyrmions subject to the gauge field originating from the magnetic quadrupole.

\textit{Current-induced  ATM skyrmion motion.---}It is well known that electrons experience an effective Lorentz force when they adiabatically pass through a skyrmion in ferromagnets, which, in return, generates a Magnus force on the skyrmion \cite{Iwasaki2013}. In antiferromagnets, the adiabatic spin transfer process plays no role in driving the skyrmion, because of the cancellation effect from different sublattices \cite{Barker2016}. However, Eq. \eqref{Eq5} indicates a finite gauge field due to the nonzero magnetic quadrupole. We thus envision an effective force when considering the adiabatic spin transfer between electrons and ATM skyrmion. To this end, we extend the Lagrangian by replacing $\partial_t$ in Eq. \eqref{Eq5} with $\partial_t+{\bf u}\cdot \nabla$, where ${\bf u}=\frac{\mu_B P}{e\gamma M_s}{\bf j}$ is the electron drift velocity, with the current density ${\bf j}$, Bohr magneton $\mu_B$, saturation magnetization $M_s$, elementary charge $e$, and polarization $P$. The energy dissipation and nonadiabatic STT can be described by the Rayleigh function
\begin{equation}\label{Eq8}
\begin{aligned}
\mathcal{R}=\frac{\alpha}{\gamma} (\partial_t {\bf l})^2+\frac{\beta}{\gamma} \partial_t {\bf l} \cdot({\bf u}\cdot\bigtriangledown) {\bf l}.\\
\end{aligned}
\end{equation}
Here, $\alpha$ and $\beta$ represent the Gilbert damping and nonadiabatic coefficient, respectively \cite{Hals2011}.
Furthermore, we assume that the moving skyrmion has a fixed shape and can be simply described by its guiding center ${\bf X}(t)$. In terms of the collective-coordinate approach \cite{Thiele,Tveten2013}, we derive the equation of motion for ATM skyrmions
\begin{equation}\label{Eq9}
\begin{aligned}
\mathop{\mathcal{M}}\limits ^{\longleftrightarrow}{\bf \dot{v}}+\mathop{\mathcal{D}}\limits ^{\leftrightarrow} (\alpha{\bf v}+\beta{\bf u})-\mathop{\mathcal{A}}\limits ^{\leftrightarrow} {\bf u}=0.\\
\end{aligned}
\end{equation}
This is the first key result of the present work. It is noted that, like its antiferromagnetic counterpart, the ATM skyrmion charge $\mathcal{C}=\frac{1}{4\pi}\int\big[ {\bf s}_A\cdot (\partial_x{\bf s}_A\times\partial_y{\bf s}_A)+{\bf s}_B\cdot (\partial_x{\bf s}_B\times\partial_y{\bf s}_B) \big] d{\bf r}$ vanishes due to the compensation of two sublattices. In this sense, a ATM skyrmion can be regarded as a ``neutral particle", and the conventional gyrotropic term $\mathcal{C}{\bf z}\times {\bf v}$ is therefore absent. Here, {$\mathcal{M}_{ij}=\frac{1}{A_0\gamma}\int {\partial_i {\bf l}\cdot \partial_j {\bf l}} d{\bf r}$} is the effective ATM skyrmion mass of a tensor form, ${\bf v}= {\dot{\bf X}}(t)$ is the skyrmion velocity, $\alpha\mathop{\mathcal{D}}\limits ^{\leftrightarrow} {\bf v}$ represents the viscous force and $\beta\mathop{\mathcal{D}}\limits ^{\leftrightarrow} {\bf u}$ denotes the drag force from the nonadiabatic STT, with the dimensionless tensor $\mathcal{D}_{ij}=\int {\partial_i {\bf l}\cdot \partial_j {\bf l}} d{\bf r}$, and $\mathop{\mathcal{A}}\limits ^{\leftrightarrow} {\bf u}$ represents the effective force from the magnetic quadrupole with the dimensionless quadrupole tensor $\mathcal{A}_{ij}=\int \partial_{j} (\mathscr{A}\cdot\partial_i {\bf l}) d{\bf r}$. Considering a steady-state motion, we derive the skyrmion velocity
\begin{equation}\label{Eq10}
\begin{aligned}
v_x&=\Gamma^{-1} \mathcal{D}_{yy}\big [(\mathcal{A}_{xx}-\beta \mathcal{D}_{xx}) u_{x} +\mathcal{A}_{yx} u_{y}\big ],\\
v_y&=\Gamma^{-1} \mathcal{D}_{xx}\big [\mathcal{A}_{xy} u_{x}+(\mathcal{A}_{yy}-\beta \mathcal{D}_{xx}) u_{y}\big ],\\
\end{aligned}
\end{equation}where $\Gamma=\alpha \mathcal{D}_{xx}\mathcal{D}_{yy}$, $u_x={|\bf u|}\cos\phi$, and $u_u={|\bf u|}\sin\phi$, with $\phi$ denoting the current flowing angle with respect to the $x-$axis. Next, we verify our theoretical predictions by full micromagnetic simulations using the \texttt{MuMax3} package \cite{SM,mumax}. The dipolar interactions are included in our simulations. 

\begin{figure}
  \centering
  \includegraphics[width=0.48\textwidth]{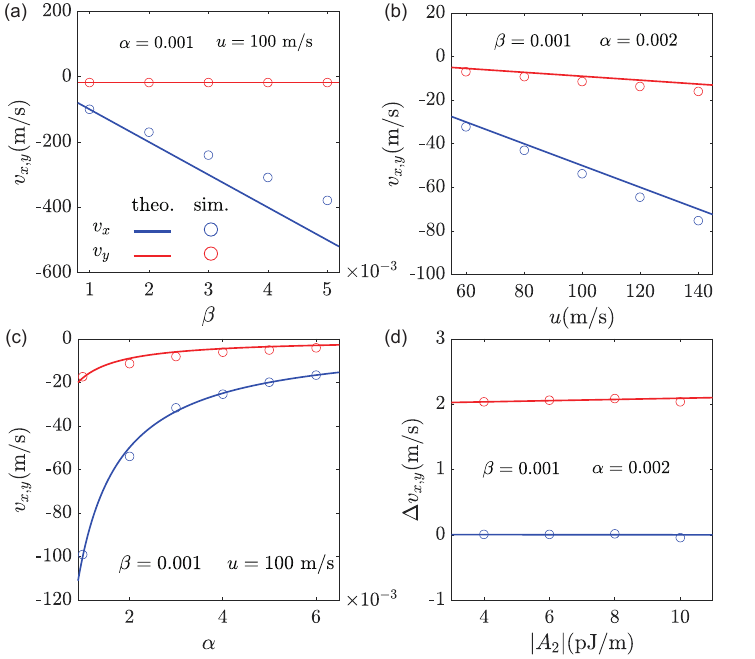}\\
  \caption{Skyrmion velocity of as a function of the non-adiabatic coefficient $\beta$ (a), charge drift velocity $u$ (b), and damping constant $\alpha$ (c) for fixed ATM exchange stiffness $A_2=10$ pJ/m and interlayer exchange constant $A_0=2$ MJ/m$^3$. (d) The difference between transverse skyrmion velocities as a function of $|A_{2}|$  with parameters: $u=100$ m/s and $A_0=2$ MJ/m$^3$.}\label{fig2}
\end{figure}

We first apply the electric current along the $x-$direction ($\phi=0$). The skyrmion velocities [Eq. \eqref{Eq10}] can then be simplified as
\begin{equation}\label{Eq11}
v_x=\frac{-\beta u_x}{\alpha}, \text{~~and~~}
v_y=\frac{\mathcal{A}_{xy}u_{x}}{\alpha \mathcal{D}_{yy}}.\\
\end{equation}
Here, we have neglected $\mathcal{A}_{xx(yy)}$, because they are high-order derivative terms compared with $\mathcal{D}_{xx(yy)}$. In this case, Eq. \eqref{Eq11} shows that the longitude (transverse) velocity of skyrmion is merely determined by the non-adiabatic (adiabatic) torque. Figures \ref{fig2}(a) and \ref{fig2}(b) plot the quantitative comparison between theoretical calculations (curves) and micromagnetic simulations (symbols) of the skyrmion velocity for different $\beta$ and $u$. Simulations results agree well with the analytical formula \eqref{Eq11} with parameters $\mathcal{A}_{xy}=-2.808\times10^{-3}$ and $\mathcal{D}_{yy}=16.1185$. In addition, we find that the skyrmion mobility is reduced by an enhanced damping parameter, as shown in Fig. \ref{fig2}(c).

\begin{figure}
  \centering
  \includegraphics[width=0.48\textwidth]{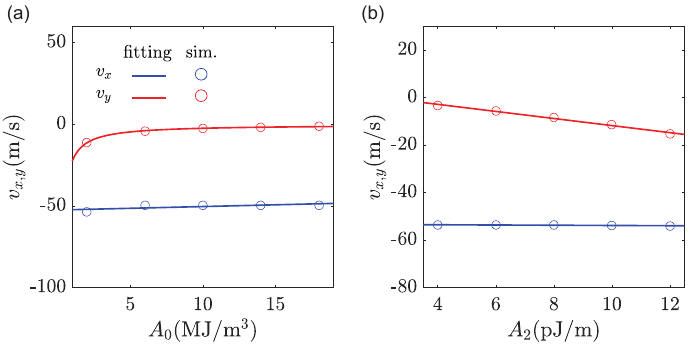}\\
  \caption{Skyrmion velocity as a function of ATM exchange stiffness (a) and interlayer exchange constant (b). Red curve represents the $1/A_{0}$ in (a) and linear fitting in (b), and blue line denotes the linear fitting in both (a) and (b). In calculations, we adopt the following parameters: $u_x=100$ m/s, $u_y=0$, $\beta=0.001$, and $\alpha=0.002$ }\label{fig3}
\end{figure}

It is worth pointing out that the sign of the gauge field $\mathscr{A}$ is determined by the exchange stiffness $A_2$ of the ATM. Consequently, one can expect an opposite transverse force when $A_2$ changes its sign. Through a symmetry analysis, we find that the driving and dissipative tensors obey the following relations: 
$\mathcal{A}_{xy,1}=-\mathcal{A}_{yx,2}$, $\mathcal{A}_{xy,2}=-\mathcal{A}_{yx,1}$, $\mathcal{D}_{xx,1}=\mathcal{D}_{yy,2}$, and $\mathcal{D}_{yy,1}=\mathcal{D}_{xx,2}$. Here, subscripts $1$ and $2$ represent positive ($J_2>J_1$) and negative ($J_1>J_2$) ATM exchange stiffness, respectively. Swapping the stiffnesses, i.e., $J_1\leftrightarrow J_2$, will generate a difference merely in the transverse velocity $\Delta v_{y}=-\Gamma^{-1}\mathcal{A}_{yx}(\mathcal{D}_{yy}-\mathcal{D}_{xx})$, which is verified by micromagnetic simulations [see Fig. \ref{fig2}(d)].

Figure \ref{fig3}(a) plots the skyrmion velocity by varying the interlayer exchange constant. It shows that $v_{y}$ monotonically decreases as the interlayer coupling increases due to the inverse proportionality between ${\bf m}_s$ and $A_0$, while $v_{x}$ stays nearly as a constant. Moreover, because the gauge field $\mathscr{A}$ is proportional to the ATM stiffness $A_{2}$, we observe a linear variation of the transverse skyrmion velocity, as shown in Fig. \ref{fig3}(b).

\begin{figure}
  \centering
  \includegraphics[width=0.48\textwidth]{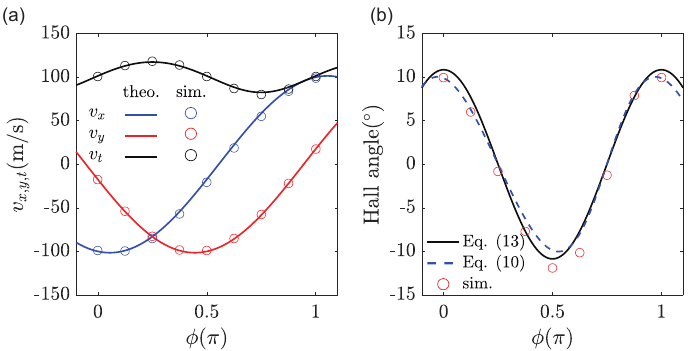}\\
  \caption{Skyrmion velocity (a) and Hall angle (b) as a function of the current direction. In calculations, we set $u=100$ m/s, $A_2=10$ pJ/m, $A_0=2$ MJ/m$^3$, $\beta=0.001$, and $\alpha=0.001$.}\label{fig4}
\end{figure}

To further understand the role of quadrupole in the ATM skyrmion Hall effect, one can rewrite the quadrupole tensor $\mathcal{A}_{ij}$ in terms of cubic harmonics
\begin{equation}\label{Eq12}
\begin{aligned}
\mathcal{A}_{x'y'}=c_{1}\cos(2\phi)\int \mathcal{Q}_{u} d{\bf r},
\end{aligned}
\end{equation}
where $c_{1}$ is the strength of the quadrupole tensor \cite{SM}, ${\bf x}'$ is the current direction, and $\mathcal{Q}_{u}=(x'^2+y'^2)$ denotes the quadrupole harmonic in cubic lattice \cite{Hayami2020}. We finally obtain the expression of the skyrmion Hall angle
\begin{equation}\label{Eq13}
\begin{aligned}
{\theta}_{s}=\arctan\big[\cos(2\phi)\mathcal{Q}_{m}\big],
\end{aligned}
\end{equation}
with $\mathcal{Q}_m=-\frac{c_1}{\beta D_{x'x'}}\int \mathcal{Q}_{u} d{\bf r}$. This is another key result of current work. In above derivations, we have neglected $\mathcal{A}_{x'x'}$ term and assumed that the N\'{e}el vector remains rotationally symmetric, which is justified by Fig. S2 \cite{SM}. One can clearly see that the skyrmion Hall effect is determined by the quadrupole harmonics and the current direction $\phi$. Figure \ref{fig4}(a) shows the magnitude of skyrmion velocities as a function of $\phi$, where the total skyrmion velocity is defined as $v_t=\sqrt{v^{2}_x+v^{2}_y}$. Notably, when the top-layer skyrmion can be mapped onto the bottom-layer one by a mirror operation about four special current direction [$\phi_m=(n+\frac{1}{2})\frac{\pi}{2}$, $n=0,1,2,3$], the skyrmion Hall effect vanishes, as evidenced by Fig. \ref{fig4}(b) which shows the $\phi$-dependence of the skyrmion Hall angle $\theta_s$. Our quadrupole model excellently predicts an anisotropic ATM skyrmion Hall effect, by comparing with micromagnetic simulations.
An intuitive understanding is as follows: When the electric current flows along the $x$ ($y$)-direction, the horizontally (vertically) stretched sublattice ferromagnetic skyrmion dominates its transport (see Fig. \ref{fig1}(a), Figs. S1 and S2 \cite{SM}). The opposite skyrmion charge will generate opposite Lorentz force on electrons, which in return induces the opposite Magnus force on the ATM skyrmion. However, when electrons flow along the mirror axis, i.e., $\phi=\phi_m$, the gauge fields from two sublattices perfectly cancel each other out, leading to a vanishing Hall effect for both the electron and ATM skyrmion. 

\textit{Discussion.---}Taking ATM skyrmion as the primary example, we have revealed the key role played by the magnetic quadrupole in the Hall transport of spin textures. This finding also applies to conventional antiferromagnetic skyrmion under an external magnetic field that can induce finite local magnetic moments. Indeed, we have observed a skyrmion Hall effect in antiferromagnets, by applying a perpendicular magnetic field (see Sec. IV in Supplemental Material \cite{SM}).

\begin{figure}
  \centering
  \includegraphics[width=0.5\textwidth]{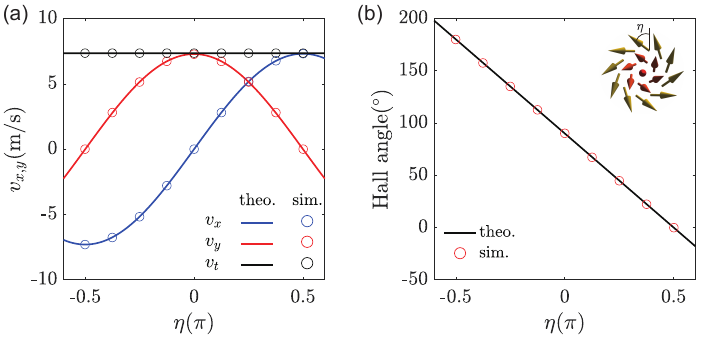}\\
  \caption{Skyrmion velocity (a) and Hall angle (b) driven by SOT as a function of its helicity $\eta$. The inset plots the helicity in the presence of hybrid DMI. In calculations, we adopt the following parameters: $A_2=10$ pJ/m, $A_0=2$ MJ/m$^3$, $\alpha=0.2$, $j=5\times 10^{9}$ A/m$^{2}$, and ${\bf p}=(1, 0, 0)$.}\label{fig5}
\end{figure}

Besides STT, the spin-orbit torque (SOT) \cite{Msiska2022,Caretta2018,Shiino2016,Zhang2} is another important knob to drive the magnetization dynamics. Unfortunately, SOT cannot induce the Hall motion for (N$\rm{\acute{e}}el$-type) ATM skyrmions. However, SOT can generate an effective force on the skyrmion helicity if one considers a hybrid DMI that supplements the original Hamiltonian \eqref{Eq1} with an extra term ${\mathcal H}_{\text{D}}=-D_{\rm B}\sum_{i,j}\big[({\bf s}_{i,j}^A\times{\bf s}_{i+1,j}^A+{\bf s}_{i,j}^B\times{\bf s}_{i+1,j}^B)\cdot{\bf x}+({\bf s}_{i,j}^A\times{\bf s}_{i,j+1}^A+{\bf s}_{i,j}^B\times{\bf s}_{i,j+1}^B)\cdot{\bf y}\big]$, with $D_{\rm B}$ being the bulk DMI constant. The skyrmion helicity is then given by $\eta=\arctan (D_{\rm B}/D_{\rm 0})$. The dissipation and SOT can be modeled by the Rayleigh function $\mathcal{R}=\frac{\alpha}{\gamma} (\partial_t {\bf l})^2+ \frac{u_{s}}{\gamma}\big [\partial_t{\bf l}\cdot({\bf l}\times {\bf p})\big ]$ \cite{Gomonay2010},
where ${\bf p}$ is the electron polarization direction, and $u_s=\gamma d j\hbar/2\mu_0 e M_s w$ denotes the strength of SOT, with the current density $j$, the reduced Planck constant $\hbar$, the vacuum permeability constant $\mu_0$, and the film thickness $w$. We then obtain the SOT-induced equation of motion for ATM skyrmions:
$\mathop{\mathcal{M}}\limits ^{\longleftrightarrow}{\bf \dot{v}}+\alpha\mathop{\mathcal{D}}\limits ^{\leftrightarrow} {\bf v}-u_{s}\mathop{\mathcal{I}}\limits ^{\leftrightarrow} {\bf p}=0$, with $\mathcal{I}_{ij}=\frac{1}{d}\int(\partial_{i}{\bf l}\times{\bf l})_{j} d{\bf r}$. Assuming the electron polarization along $x-$direction, one gets the steady skyrmion velocities $v_x=\mathcal{I}_{xx}u_{s}/(\alpha D_{xx})$ and $v_y=\mathcal{I}_{yx}u_{s}/(\alpha D_{yy})$. Micromagnetic simulations agree well with the above formulas [Fig. \ref{fig5}(a)]. It is noted that, unlike the magnetic moment ${\bf m}_s$, the staggered parameter $\bf l$ is isotropic (see Fig. S2 \cite{SM}). We therefore observe an $\eta-$independent total velocity of skyrmion, as shown by the black line in Fig. \ref{fig5}(a). By assuming an $\eta-$independent $\mathcal{D}_{ij}$ and $360 ^\circ$ domain-wall solution of the skyrmion profile \cite{Wang2018}, one can analytically derive the skyrmion Hall angle $\theta_s=\frac{\pi}{2}-\eta$, consistent with micromagnetic simulations, see Fig. \ref{fig5}(b). Importantly, it indicates that the magnetic quadrupole does not play a role when driven by SOT. This underscores the importance of adiabatic phase that is a key ingredient of the AC effect.

\textit{Conclusion.---}To summarize, we predicted an emerging skyrmion Hall effect in altermagnets, by establishing an isomorphism between the dynamics of ATM skyrmion and AC effect. We showed that the local magnetization in ATM skyrmion constitutes a magnetic quadrupole that leads to an anisotropic skyrmion Hall effect when driven by STT. A sign change of the Hall angle was identified when the ATM exchange stiffness is swapped. Due to the absence of adiabatic transport, SOT cannot induce the ATM skyrmion Hall effect, unless it couples with the skrymion helicity in the presence of hybird DMI. Our findings reveal the hidden gauge field concealed in the magnetic quadrupole of ATM skyrmion and would significantly advance the understanding of the Hall effect of neutral quasiparticles beyond skyrmions.

\begin{acknowledgments}
 This work was funded by the National Key R$\&$D Program under Contract No. 2022YFA1402802 and the National Natural Science Foundation of China (NSFC) (Grants No. 12374103 and No. 12074057).
\end{acknowledgments}

Z. J. and Z. Z. contributed equally.

\end{document}